# SIMULATION FRAMEWORK FOR MODELING LARGE-SCALE DISTRIBUTED SYSTEMS

Dobre Ciprian Mihai *, Cristea Valentin *, Iosif C. Legrand **

\* "Politehnica" University of Bucharest
\*\* California Institute of Technology

Simulation has become the evaluation method of choice for many areas of distributing computing research. However, most existing simulation packages have several limitations on the size and complexity of the system being modeled. Fine grained simulation of complex systems such as Grids requires high computational effort which can only be obtained by using an underlying distributed architecture. We are proposing a new distributed simulation system that has the advantage of being able to model very complex distributed systems while hiding the computational effort from the end-user.

## 1. INTRODUCTION

Optimizing the use of Grid resources is crucial, and simulation is the key to evaluate potential optimization strategies. Simulation has become the evaluation method of choice for many areas of distributing computing research. However, most existing simulation packages have several limitations on the size and complexity of the system being modeled. Simulated systems of just a few thousands computing elements and a few thousands data flows will quickly exhaust the computing resources in any reasonable sized computer workstation. Many times it is important to simulate Grid resources as realistically as possible before they are used on real Grids.

One way of copping up with the increasingly power demand coming from the simulation scenarios nowadays is to make use of more processor units, running on different architectures and dispersed around a larger area, in other words one way of keeping up with the simulating scenarios is to distribute the simulation application. Distributed discrete event simulation techniques aim at an acceleration of the execution of a self-contained simulation models by spatial decomposition and the concurrent simulation of sub-models, using the concept of logical process, each one executing on a dedicated node of a (closed) multiprocessor system. When simulating large scale models such as Grid systems the distributed event simulation is the alternative to use.

## 2. RELATED WORK

Even with more then 25 years of experience behind, having a broad body of theory and methods being developed in all this time, the field of distributed discrete event simulation failed in generating general acceptance in simulation practice. The field has suffered from simulations and industrials being reserved on the potential gains of these methods, considering the complexity of development and implementation efforts. Nevertheless, a series of pioneer projects in this field opened up new frontiers for research, such as GTW (Georgia Tech), CPSim (Boyan Tech Inc), TWOS (Jet Propulsion Laboratory), WARPED (University of Cincinnati) or APOSTLE (DRA Malvern).

For simulating Grid systems a number of simulation projects already exists and were successfully used in modeling various Grid related technologies. Projects such as MONARC, Bricks, ChicagoSim, GridSim, SimGrid or GridNet already proved their values, but they all suffer from limitations on the size of the distributed system being modeled.

With the ever-increasing need for validating various Grid systems today more then ever there is a need for a simulation framework capable of combining the advantages of the distributed discrete event simulators with the simulation models already validated by the Grid simulators. Such a simulation framework would make it possible to verify virtually unlimited in complexity Grid systems scenarios.

## 3. THE PARALLEL SIMULATION FRAMEWORK

The proposed distributed simulation framework is based on the experience gained in the implementation of a parallel simulation framework. The simulation project, MONARC, was chosen because its simulation model, based on the concept of regional centers, was successfully used in constructing a very large number of complex simulation scenarios. The components of a regional center are represented in figure1.

The aim of the MONARC project was to provide a realistic simulation of large distributed computing systems and to offer a flexible and dynamic environment to evaluate the performance of a range of possible data processing architectures. To achieve this purpose, the simulator provides the mechanisms to describe concurrent network traffic and to evaluate different strategies in data replication and job scheduling procedures.

The simulation model proposed in MONARC was already been tested on a large variety of scenarios. The validity of the simulation model is what makes it an excellent simulation instrument that constitutes the basis for implementing the new distributed simulation framework.

### 3.1 A simulation study for T0/T1 data replication and production analysis.

One of the most impressive simulation studies involving MONARC tested the behavior of the two largest experiments at CERN, namely CMS and ATLAS. The general concept developed by these experiments is a hierarchy of distributed Regional Centers working in close coordination with the main center located at CERN. This simulation study followed this concept and described several major activities; mainly the data transfer on WAN between the T0 (CERN) and a number of several T1 Regional Centers. The obtained results actually have shown that for the link connecting CERN to US a minimum 10 Gbps bandwidth was necessary and also proved

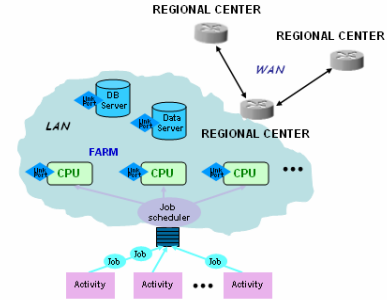

Fig. 1.The regional center.

the need for use of a data replication mechanism in the connecting nodes.

What was noticed during this experiment is the time needed to actual complete the complex simulation runs involved in this experiment. The actual values are presented in figure 2.

The data were obtained in the simulations done using different values for the entry set (in this case the available bandwidth between Europe and US). For all the runs of the simulation scenario a unique machine was used, having the following hardware configuration: 2 Intel Xenon CPUs at 2.4 GHz and 4 GB RAM.

There are two reasons why the time needed to finish the simulation runs tends to behave as an exponential function as seen in the graph above. The first one is directly related to the simulation events processed by each task (active object) involved in the simulation model.

The increase in the bandwidth is proportional with the number of simulated messages sent throughout the network. When there is enough bandwidth available one simulated message starts being transmitted and it is totally sent to its destination in a reasonable amount of time. When there is not enough bandwidth in the simulated system the time needed to effectively transmit a simulated message is higher, meaning that another simulated message that is transmitted at a later time will interrupt the first one with a higher rate of probability. Because the interrupt mechanism is based on simulation events we observed an increase in the number of events that each active job has to deal with. So processing the events is one reason why the time needed to complete the simulation varies so much.

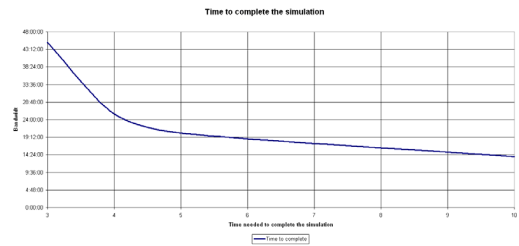

Fig. 2. Effective time needed to complete the simulation runs using different parameters.

One other fact observed during the simulation runs is that in fact a larger number of messages lead to an increase in the used physical memory of the simulator. This has two direct consequences. During the simulation runs the physical load of the machine increased from the simulation done using a higher value for the available bandwidth to the same experiment done using a lower value for the bandwidth. So first the processing unit of the physical host acted as a bottleneck to the simulation. Then the consumed memory of the application also increased from the simulation run done using a higher value for the available bandwidth to the same experiment done using a lower value for the bandwidth. So the physical memory available on the physical machine tends also to act as a bottleneck to the parallel simulation application.

## 4. THE DISTRIBUTED SIMULATION SYSTEM

In the design and implementation of a distributed simulation framework several aspects have to be considered: computation decomposition, process allocation and synchronization. The computation processes should be distributed across the processors in order to balance the load. The proposed strategy to use is based on dynamic decomposition, meaning the scheduling algorithm dynamically assign processes to processors. The decision about the adequate allocation technique strongly depends on the hardware platform and the characteristic features of the simulation study considered and are linked together with a monitoring framework in order to correctly balance the computational load across the nodes of the distributed system.

When implementing a distributed discrete event simulation framework one needs to consider the aspect of ensuring the causal constraints between the simulation events. This corresponds to finding an adequate synchronization mechanism. In our proposal we adopt a conservative simulation synchronization mechanism. The optimistic synchronization protocols pose complex problems such as state saving that are not acceptable for the distributed simulation models proposed around the concept of Regional Centers. It would be difficult to construct a distributed simulation system using optimistic protocols because the state of the system is hard, if not impossible, to be determined. The object-oriented approach of describing various simulated components makes the determination of the set of parameters that fully determine the state of the system near impossible. Then the problem of keeping the state saving synchronized between the simulation components is another difficult issue to deal with. By using a conservative synchronization mechanism we are aiming at providing simplicity and generality to the distributed simulation system.

In case of its architecture the simulation framework consists of a set of simulation agents distributed among the physical nodes of the underlying distributed system where the application is deployed. A simulation model is distributed and executed among a set of the entire collection of running agents in the system, where the decision on which the set is chose is described next. Each simulation agent executes a set of logical processes. Each logical process operates as an active object, meaning it has an execution thread, program counter, stack, etc, and is able to execute simulation events. The logical processes running on a simulation agent may belong to the same simulation run spread over the simulation system or may belong to different simulation runs.

The problem of dynamic lookup of the simulation agents across the network is addressed by a set of lookup services based on Jini technology. The distributed simulation system is also linked to a dynamic monitoring service in order to implement the scheduling algorithm.

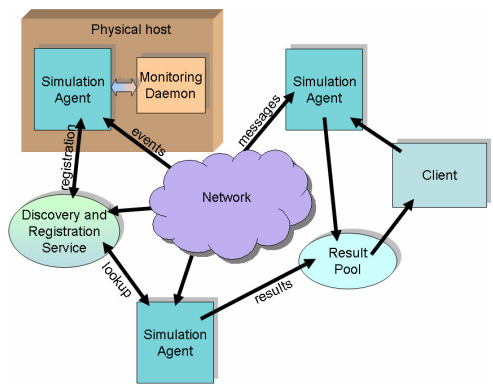

Fig. 3. The distributed simulation system.

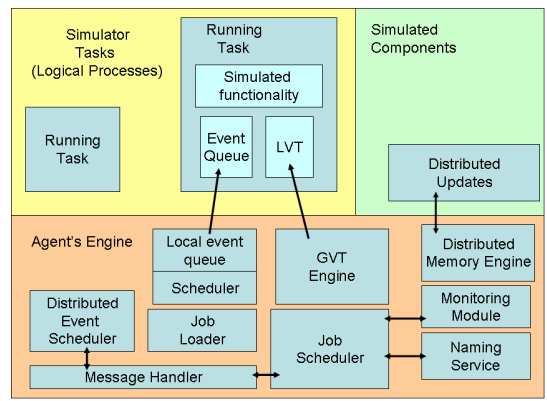

Fig. 4. The Simulation Engine.

The proposed distributed infrastructure is best described in figure 3. Figure 4 presents the functionality of the simulation agent.

For the design of the Simulation Agent we are adopting a layered architecture. On the bottom we have the Engine component. This element is responsible for the appropriate dispersal of simulation events to the suitable logical processes, for scheduling new simulation jobs properly, for bridging the components of the simulation model all together, and for keeping the simulation in a causal relationship consistency.

On top of the engine are the logical processes that run locally on the simulation agent and the simulation components that are part of the executed simulation models. The engine is responsible with distributing the events created during the execution of logical processes and with ensuring the consistency of the global virtual time. The logical processes are grouped based on the simulation context and the addressing scheme.

The user deploys the simulation scenario on one of the deployed simulation agents. Completely transparent the agent will distribute the computing effort among other existing simulation agents dynamically found using the lookup service. When a new simulation job is created in the simulation scenario, the scheduler will deploy it for execution on an appropriate simulation agent. Whenever possible the engine will try to reuse the existing local logical processes, but the result of a newly created simulation job can also be the creation of a new logical process.

*4.1 The Scheduling Algorithm.*

The scheduling algorithm uses performance monitoring to optimize the execution time of each simulation job. Linking the distributed simulation application with a monitoring system represents a premiere. By using this approach the job scheduler is able to load-balance the execution of the active processes between the nodes of the distributed infrastructure. The monitoring system we propose using is LISA (Localhost Information Service Agent). LISA is an easy-to-use monitoring system that can help optimizing other applications and offers a variety of of-the-shelf monitoring module that can be dynamically loaded inside the system.

After studying several distributed simulation systems we found that there are a number of factors that affects the overall performance of the simulation system. First of all the limitations of the workstations that compose the distributed system are an important decisional factor to consider. By assigning simulation jobs to be executed on slow workstation all other simulation jobs are affected. This happens because of the need to maintain causal consistency between the executing jobs. Consider for example the case when we keep jobs blocked because their execution can not progress until the local virtual time computed on slower workstation is synchronized. The latency of the network infrastructure of the distributed factor is also a limiting factor to consider. The distributed system is build around the concept of message passing, so the faster the messages reach there destinations the better the performance shall be. The final aspect to consider is the simulation model itself. By assigning jobs to be executed on the same fastest workstation can prove to be also a limitation. This means that sometimes it is best to schedule two simulation jobs for execution on different workstations that on the same one. Generally, at any given moment of time, the CPU unit of the workstation is able to handle only one process, so if the context switching between processes takes longer that would take for the network message passing then its best to distribute the simulation tasks.

The algorithm of the simulation job scheduler is based on the performance value of each simulation agent. Each simulation agent publishes a performance value that can be used by any other existing simulation agent. This performance value takes into consideration the load of the physical workstation where the agent is running (cpu load, available memory, etc.), the load of the network (distances between agents, round-trip-time, available bandwidth, etc.) and also the load of the agents (number of logical processes already executing on top of the simulation agent, what components are already duplicated locally, etc.).

At any given moment a logical process can decide to schedule the execution of a new simulation job. The simulation agent accesses the performance values of all other simulation nodes. Using the performance values and the topology of the distributed system the agent computes an undirected graph. The graph is weighted and complete, and we associate to any edge a value computed as the arithmetic mean between the performance values of the two connecting vertices of the edge. On this graph we compute next the shortest paths between any two vertices of the graph. After this step for each node we obtain a list of values that represents the values of the shortest paths between that node and the rest of the other nodes. From this list we remove the values of the shortest paths between that node and nodes that are not yet participating in the simulation run. The remaining values are then used to obtain a new performance value. This new performance value represents the arithmetic mean of the remaining values. By sorting the values the node on top of the list is the preferred node on which we can schedule the execution of the new simulation job.

This algorithm is efficient because it takes into consideration a full range of parameters, starting from the physical running parameters of the workstations belonging to the distributed system and ending with the performance values of the network links connecting the nodes, as already described. What is even more interesting to note is that it tries to group the logical processes belonging to the same simulation run into a minimum cluster graph of nodes, limiting in this way the number of messages that are exchanged between the logical processes during the simulation run. Also, the distributed components of the simulation model are grouped based on proximity, though simplifying the process of synchronization between them.

*4.2 Simulation Components.*

When running Grid simulations we need to generate synthetic Grid platforms, including two main elements – the network resources and the compute resources – with, in both cases, models for „background" resource utilization and availability. One distinguishing features of Grid platforms when compared to traditional parallel computing platform is the complexity of the system. This complexity leads to heterogeneity of bandwidths and of round-trip times as well as complex bandwidth sharing among competing network connections and the heterogeneity of the computing resources.

To correctly describe a large variety of distributed systems architectures the system assumes that the simulation model consists of a number of simulation components, such as CPU units, database servers, network components, farms and regional centers. All these components are actual Java objects, described by their implementation, and the state of their attributes at any moment of time. The state of the simulation jobs is a function of the current states of the simulation components. For example a processing job depends on the values of the processing power and available memory of the simulated CPU unit on which it is executed. The proposed solution for implementing the distributed simulation application is to circulate the simulation components objects among the computing nodes.

This approach can prove to be optimum in terms of overall performance of the simulation application. By using replicas of the same component objects distributed among computing nodes involved in the simulation we are not imposing a limitation to where a logical processes will be executed. In fact, as already described, the scheduler algorithm does not consider any such limitation. As an example let us consider the case of two processing jobs. If we would limit the model to force that the CPU unit object used by the two simulation jobs reside on only one workstation we would face two problems. First it would be difficult to schedule the execution of the two jobs on two separate workstations. That particular workstation would become a processing bottleneck in the simulation while other nodes of the distributed systems would not be used much. So we would break the principle of load balancing between the nodes involved in the simulation. What happens when three or more jobs are scheduled to be executed on the same CPU unit, or even more? In essence, we would use only as many workstation as the number of CPU units declared in the simulation model. This is why using replicas of the same objects between the processing nodes of the distributed system is a better alternative in terms of performance.

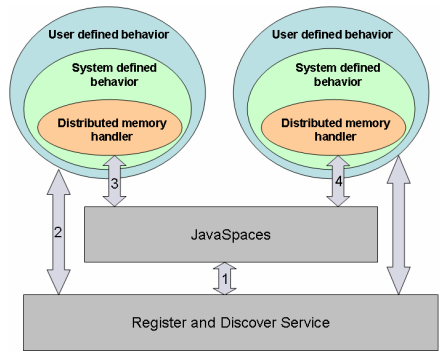

Fig.5.Implementation of the simulation components.

The simulation components of the simulation model are constructed starting from basic implementations of them. The basic implementations of the components are defined from the beginning inside the distributed application. Each basic component implementation provides two aspects to the simulation model. First it provides methods describing the basic functionality of that specific simulation component. Besides that it provides means to synchronize the values of the fields defining the component state among other such components. One should think of this approach as replicated distributed objects. We are interesting in two aspects here: how the objects discover each others and then how the values of their states are kept in a synchronized state. The proposed approach is represented in the figure5.

The state consistency of various replicas of the same objects is imposed using a distributed memory implementation based on JavaSpaces. JavaSpaces technology provides great power and offers support for advanced features such as persistant storage, secure transactions and event notification. The distributed objects are based on a reactive style of programming, based on Jini's distributed event model.

Each simulated component extends a basic component class that offers methods for working with the distributed memory space. Because the distributed application is object-oriented all extended components can define more complex operations using the state parameters and the basic behavior methods already offered. This approach is needed in order to define complex simulation model. The advantages of this approach are: the user can easily define new simulation components using the methods offered by the basic implementations (one could see the offered methods as "building blocks" for more complex component specifications); there is no limitation imposed to the user by this model, he can even define new components starting from the basic class; parameters are updated transparent from the user – the user does not need to know what happens with the updates of the parameters, he does not care that the parameters are distributed among replicas of the same component object, he does not even have to know that his simulated component will not reside in a single context of any process.

The distributed simulation framework provides a set of basic components. One such component is the data model. The data model should provide a realistic mapping of a DBMS, and, at the same time, allow an efficient way to describe very large database systems with a huge number of objects. For simulating the databases, two main entities used to store data will be modeled: the database server and the mass storage center. The database server stores the data on disk drives, while the mass storage center uses tape drives. The users of the distributed system can interact with both these entities, but the simulation framework also provides an algorithm that automatically moves the data from a database server to the mass storage server(s) when the first one is out of storage space.

In regard to the network model, the simulation program offers the possibility of simulating data traffic for different protocols on both LAN and WAN. The proposed approach used to simulate the data traffic is again based on the "interrupt" scheme.

Beside the basic components described above the distributed simulation framework should provide a series of components specific to Grid simulations, such as metadata catalog, analysis jobs or a distributed job scheduler. These components were first proposed by MONARC and have proven to be more then necessarily in order to construct complex simulation scenarios.

Other components of the proposed simulation framework are represented by the client and the result pool. The client provides means through which a simulation run is started and also provides a visual reference to the user on the current state of the simulation being executed. The implementation of the client is also based on Java technology and offers transparent access to the deployed simulation agents. The client could be a Java WebStart application that can connect to any simulation agent, send commands and receive simulation results. The result pool is the component that runs inside the client and is responsible with their interpretation. The pool can also save results locally. This action can have two main advantages. First the simulation can be evaluated at a later moment of time without rerunning the complete model. Then the simulation results can be used as input for another simulation run.

*4.3 The synchronization mechanism.*

The simulated events are in a causal relationship that must be considered. It is very important to impose the causality constraints in order to ensure the correctness of the simulation. Violating causality constraints means that the future can affect the past. This can result in anomalous behavior and consequently incorrect simulation. It is the responsibility of the synchronization mechanism to ensure proper and correct interactions among the logical processes.

The advantages of using a conservative approach instead of optimistic protocols were already presented. The characteristic of a conservative implementation is that it absolutely avoids the occurrence of causality violations. This is done by only allowing an event to be processed if it can be guaranteed that no message with a lower timestamp will be received in the future. This means that our implementation does not need to save the state of the simulated model.

A vast majority of the conservative simulation algorithms are based on the CMB algorithm (Misra, 1986). What is interesting about the CMB algorithm is that several important observations made in the algorithm epitomize the fundamentals of conservative parallel simulation algorithms. The proposed implementation of the conservative simulation algorithm is also based on the concepts proposed by the CMB algorithm.

A simulation event is always created by a logical process and is destined to the same or other logical process. A simulation event includes information regarding the identifiers of the source logical process and of the destination logical process. In the implementation the simulation agent keeps an event queue with the events that must be processed by the local running logical processes. A scheduler component is responsible with managing the running logical processes on any simulation agent.

For the creation of logical processes a pool of worker threads is used. This eliminates the overhead caused by creating new threads and destroying them. When a new logical process is created, the scheduler takes from the pool a worker thread. At a moment of time, a logical process can be in one of five possible states: *created*, *ready*, *running*, *waiting* and *finished*. A new logical process is in the *created* state until the scheduler finds in the pool a worker thread that can execute it; then, the logical process moves to the

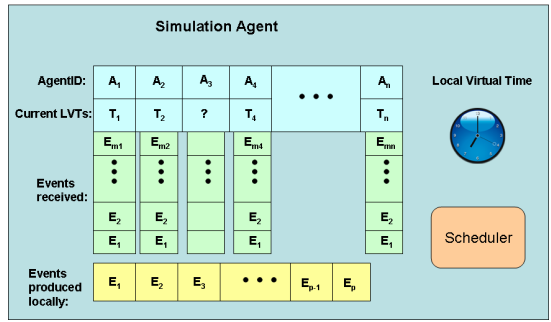

Fig.6. Structures used by the event scheduler.

*ready* state. The scheduler will let all the *ready* logical processes to run (and set their state to *running*) after it finishes processing the events from the current simulation step. When a logical process must stop its execution (for example, if it has to wait for an event), it moves to the *waiting* state.

Because the simulation model is composed of more then one simulation agents concurrently executing the scheduler must check that there are no other simulations agents in the system having the current LVT value (local virtual time) lower then the minimum timestamp in the event queue. This condition guarantees that in fact no other logical process involved in the simulation will produce new simulation events that should have been processed in the past.

In order to implement this checking the event scheduler uses a queue structure that holds the values of the virtual clocks last known for each of the other simulation agents involved in the simulation run.

In this scenario the events received from logical processes running on other simulation agents involved in the running simulation are kept in separate queues as in figure 6. One separate queue is used to keep the events produced by the local logical processes. The LVT queue is used in order to keep track of current dependencies between the values of LVT on various running nodes of the distributed simulation.

In each step the scheduler inspects the queues and decides on the events having the lowest timestamp value. The timestamp value is then compared with the known LVT values of different simulation agents. The simulation agents for whom the known LVT values are higher or equal with the value of the timestamp are guarantying that will not produce events with lower timestamps in the future, thus the causal consistency between current simulation agent and that particular simulation agent is guaranteed.

In some event queues is possible that we do not have any event. This means that for some simulation agents is possible that we do not know the current value of their local virtual clock. When this happens we can request that agent to send us its current LVT value. This means that, if the scheduler does not know the value of the current LVT value for a particular simulation agent, it will send a message containing the value of the current logical clock. The remote simulation agent can then respond back when it decides that from its point of view it is safe for the local scheduler to continue processing with the next event.

The current LVT queue is modified with the following conditions. If the simulation agent receives a new event from a remote agent it will inspect it. If it is not a "start new job" type of event and 1) the timestamp of the event is lower that the current LVT value known in the LVT queue for that particular simulation agent the LVT queue or 2) the current LVT value of the that particular simulation agent is unknown, then the LVT queue is updated. The event is then added to the queue of events corresponding to the simulation agent.

If the simulation agent receives a message containing a request for the current LVT value it will inspect it. If the LVT value contained in the message is lower then the current known LVT value or if the current LVT value for that particular agent is unknown then the LVT queue is updated.

In each step the scheduler can decide that it can not continue because the LVT values known for other simulation agents are lower that the current timestamp value. For each of these simulation agents the scheduler will execute: send a message and block until the value of the LVT value is higher then the value of the current timestamp. Whenever the LVT queue is updated the scheduler is unblocked and it will check again this condition.

The scheduler algorithm is in fact an adaptation of the null messages by demand algorithm (Fersca, 1995).

Interesting to note about this algorithm is that the number of messages exchanged between simulation agents is kept at a minimum level. Instead of synchronizing logical processes we are synchronizing the distributed simulation agents altogether.

Only one message is used to ask for the current value of the remote virtual time and also to send the local current value of the logical clock. Using this algorithm it is possible that the scheduler will process more than one event before sending out other synchronization messages. Because of these facts the proposed algorithm will prove to be much faster then any other conservative simulation algorithms known today.

*4.4 Optimizing the distributed simulation system.*

For executing more then one simulation run in parallel using the deployed simulation agents the solution is based on the concept of context. Each simulation agent will execute a set of event schedulers in parallel, each one running the algorithm described in the previous section. The logical processes are grouped in sets based on the simulation run to which they belong. The same naming service can be used to deploy this scenario, meaning that when executing a simulation run the event scheduler for example will only discover those logical processes that are executing jobs belonging to the same simulation run. Even if the distributed simulation infrastructure is the same, no object involved in one simulation run will affect other simulation objects involved in other simulation runs. The concept of context is best illustrated in figure 9.

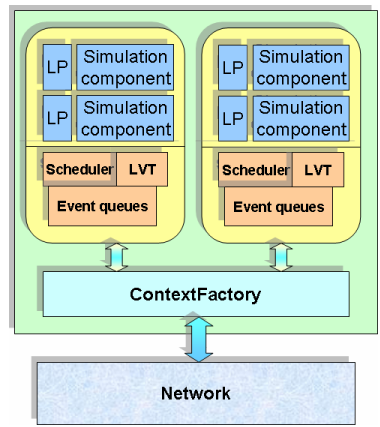

Fig. 9. The context approach of the distributed simulation system.

The context factory is responsible with correctly delivering simulation event to simulation contexts, as it is also responsible with dynamically creating or destroying new simulation contexts when a simulation run is executed or when a simulation run finishes. It links together the simulation agents involved at a given moment of time in a simulation run and keeps a clear separation between the existing simulation contexts.

Using this approach the deployed distributed agents are better used at their real capacity. By allowing the simulation agents to execute more then one simulation run in parallel at any moment of time the proposed implementation can prove even more appealing to end-user. The application permits not only the distribution of the processes involved in executing a simulation model, but also the distribution of separate simulation runs on different computing resources of the underlying platforms.

## 5. CONCLUSIONS

In terms of optimization the proposed simulation system guarantees some interesting facts. By using dynamic registration and discovery the simulation agents are receptive to the dynamic of the system in the sense that they can cope with the different types of failures that can occur inside the system. The load-balancing of the system is also assured in the way that the proposed unique scheduling algorithm is ensuring best-effort distribution of the computational processing effort. By using Java technology the user does not need to be aware of the underlying platform executing the simulation at all, he is shielded from the implementation details of the actual execution of the simulated model. The proposed distributed simulation system offers an object-oriented programming framework to the user, meaning the development time of the application is much reduced. The system provides a set of basic components that can be used to deploy a large variety of simulation scenarios off-the-shelf. In the same time the system permits the modeling of other simulation components easily. The dynamic of the simulation jobs is hidden from the user and the behavior of the modeled system does not need to be known prior to the actual execution of the simulation, those giving the user complete power in constructing various simulation scenarios.

These are only a few of the advantages that the proposed distributed simulation system provides. It is meant to be a simulation framework capable of easily model complex systems and that offers transparency of the underlying distributed system on top of which it is deployed.